\documentclass[letterpaper,english,reprint, aps]{revtex4-1}
\pdfoutput=1
\usepackage[T1]{fontenc}
\usepackage[latin9]{inputenc}
\usepackage{booktabs}
\usepackage{amsmath}
\usepackage{amssymb}
\usepackage{graphicx}
\usepackage{subscript}

\makeatletter

\pdfpageheight\paperheight
\pdfpagewidth\paperwidth

\providecommand{\tabularnewline}{\\}

\makeatother

\usepackage{babel}
\begin{document}
\title{A perturbative approach to the polaron shift of excitons in transition
metal dichalcogeniedes}
\author{J. C. G. Henriques$^{1,2}$ and N. M. R. Peres$^{1,2}$}
\address{$^{1}$Department and Centre of Physics, and QuantaLab, University
of Minho, Campus of Gualtar, 4710-057, Braga, Portugal}
\address{$^{2}$International Iberian Nanotechnology Laboratory (INL), Av. Mestre
José Veiga, 4715-330, Braga, Portugal}
\begin{abstract}
In this paper we study the phonon's effect on the position of the
1s excitonic resonance of the fundamental absorption transition line
in two-dimensional transition metal dichalcogenides. We apply our
theory to WS$_{2}$a two-dimensional material where the shift in absorption
peak position has been measured as a function of temperature. The
theory is composed of two ingredients only: i) the effect of longitudinal
optical phonons on the absorption peak position, which we describe
with second order perturbation theory; ii) the effect of phonons on
the value of the single particle energy gap, which we describe with
the Huang Rhys model. Our results show an excellent agreement with
the experimentally measured shift of the absorption peak with the
temperature.
\end{abstract}
\maketitle
One of the most prominent and studied types of two-dimensional materials
are transition metal dichalcogeniedes (TMDs) \citep{wang2012electronics}.
These, often semi-conducting, materials present remarkable electronic
and optical properties, intrinsically related with their excitonic
response. An exciton is a quasi-particle corresponding to a bound
electron-hole pair interacting via a Coulomb-like potential. Due to
the reduced dielectric screening in two-dimensional materials, these
quasi-particles are more tightly bound, and thus more stable, than
their three-dimensional analogues. Phonons, the quanta of atomic vibrational
energy, are known to have a significant impact on the optical properties
of TMDs, especially due to their interaction with excitons \citep{li_excitonphonon_2021,li2018optical,wang2017influence,shree2018observation}.
The exciton-phonon coupling influences both the line width and the
peak position of the different absorption resonances associated with
the optically active excitonic states in TMDs. Indeed, this effect
has been reported recently in Ref.\citep{raja2018enhancement}, where
it was shown that the $1s$ excitonic peak was red-shifted as the
temperature increased, accompanied by a concomitant increased line
width of the resonance. The coupling of phonons to excitons also affects
their radiative lifetime, and allows the access to optically dark
states via inter-valley scattering channels \citep{christiansen2017phonon}.
Studies on the temperature dependence of the optical properties of
these materials are highly relevant to accurately predict their applicability
in different technological applications which are required to work
at room temperature.

The problem of electron-phonon interaction is by no means a simple
one, giving rise to, for example, phonon-mediated supercondutivity
and the polaron problem, an electron dressed with a cloud of phonons.
While the former problem can be dealt with an approximate canonical
transformation, the latter one is, in general, non-perturbative. However,
to address the effect of phonon's on the position peak of the absorption
resonance a perturbative approach, up to seconder in the electron-phonon
interaction, suffices. However, as derived from traditional perturbation
theory, we end up with a sum over all states of the non-interacting
problem, whose effective summation is out of reach simply because
it requires all the eigenstates of the non-interacting system, which
may not be known. Even in the cases where they are known, the integrals
are of insurmountable difficulty. Therefore, we follow a different
path for circumventing the sum over states. We use the Dalgarno-Lewis
approach \citep{dalgarno1955exact} which shifts the sum over states
problem to the solution of an non-homogeneous differential equation.
In this procedure, only one eigenstate of the unperturbed theory is
required together with the solution of the aforementioned differential
equation.

In this paper we consider a two dimensionalTMD whose electrons and
holes, interacting via a Coulomb-like potential, may give rise to
exciton formation. In order to study the effect of temperature on
the excitonic properties, we use a similar model to the one employed
in Ref. \citep{de1973ground}, where a Frohlich-like Hamiltonian \citep{frohlich1954electrons}was
used to characterize the interaction of optical phonons with electrons
and holes in polar crystals. Contrary to Ref. \citep{de1973ground},
where 3D systems were considered, we will focus on excitons on 2D
materials, leading to a difference in the form of the interaction
term \citep{devreese2012physics}. Moreover, contrary to the aforementioned
work where only the case of $T=0$ K was considered, our calculations
cover any temperature value. The Hamiltonian of the considered system
in the center of mass frame of the electron-hole pair reads:
\begin{equation}
\mathcal{H}=H_{0}+H_{1}+H_{2}+H_{3}+H_{4},
\end{equation}
where
\begin{align*}
H_{0} & =\frac{p^{2}}{2\mu}+V(r),\quad H_{1}=\sum_{\mathbf{q}}\hbar\omega_{\mathbf{q}}a_{\mathbf{q}}^{\dagger}a_{\mathbf{q}}\\
H_{2} & =-\frac{U}{A^{1/2}}\sum_{\mathbf{q}}\frac{i}{\sqrt{q}}a_{\mathbf{q}}e^{i\mathbf{q}\cdot\left(m_{h}/M\right)\mathbf{r}}+h.c.\\
H_{3} & =\frac{U}{A^{1/2}}\sum_{\mathbf{q}}\frac{i}{\sqrt{q}}a_{\mathbf{q}}e^{-i\mathbf{q}\cdot\left(m_{e}/M\right)\mathbf{r}}+h.c.\\
H_{4} & =\frac{1}{2M}\left(\mathbf{K-\sum_{\mathbf{q}}\hbar\mathbf{q}}a_{\mathbf{q}}^{\dagger}a_{\mathbf{q}}\right)^{2}.
\end{align*}
The term $H_{0}$ is the Hamiltonian of the exciton, with $m_{e/h}$
the electron/hole effective mass, $\mu^{-1}=m_{e}^{-1}+m_{h}^{-1}$
the reduced mass of the electron-hole pair, $p$ their relative momentum,
$\mathbf{r}$ their relative position vector and $V(r)$ the electron-hole
interaction potential which we model using the Rytova-Keldysh potential
\citep{keldysh1979coulomb,rytova1967}
\begin{equation}
V(r)=-\frac{e^{2}}{4\pi\epsilon_{0}}\frac{\pi}{2r_{0}}\left[\mathbf{H}_{0}\left(\frac{\kappa r}{r_{0}}\right)-Y_{0}\left(\frac{\kappa r}{r_{0}}\right)\right],
\end{equation}
where $e$ is the elementary charge, $\epsilon_{0}$ is the vacuum
permittivity, $\kappa$ is the mean dielectric constant of the media
above and below the TMD monolayer, $r_{0}$ is a material-dependent
screening length (which is microscopically related to the material's
polarizability), $\mathbf{H}_{0}$ is the Struve function of the first
kind and $Y_{0}$ the Bessel function of the second kind, both of
order zero. In the total Hamiltonian, the term $H_{1}$ describes
the thermally excited phonons and$a_{\mathbf{q}}^{\dagger}$/$a_{\mathbf{q}}$
refers to the creation/annihilation operator of a phonon with momentum
$\mathbf{q}$ and energy $\hbar\omega_{\mathbf{q}}$. In this term
we only consider the contribution originating from longitudinal-optical
(LO) phonons. Also, we will consider that the energy $\hbar\omega_{\mathbf{q}}$
is independent of momentum and equal to a constant value $\hbar\omega_{{\rm LO}}$
when numerical results are computed\@. The terms $H_{2}$ and $H_{3}$
correspond to the interaction between the phonons and the electrons
and holes, $A$ is the area of the 2D monolayer, $M=m_{e}+m_{h}$
and $U$ is the coupling potential defined as \citep{devreese2012physics}
\begin{equation}
U=\hbar\omega_{{\rm LO}}\left(\sqrt{2}\pi\alpha\right)^{1/2}\left(\frac{\hbar}{m_{0}\omega_{{\rm LO}}}\right)^{1/4},
\end{equation}
with $m_{0}$ the bare electron mass, $\alpha$ a dimensional coupling
constant, which we will consider as a fitting parameter, but whose
typical value is between 2-5 \citep{devreese2012physics}. Finally,
the term $H_{4}$ depends on the center of mass momentum $\mathbf{K}$.
It is not in general expected that a term which depends on the center
of mass momentum will play a significant role in the system's internal
dynamics, thus hereinafter, just like in Ref. \citep{de1973ground},
we neglect its contribution to $\mathcal{H}$, that is
\begin{equation}
\mathcal{H}\approx H_{0}+H_{1}+H_{2}+H_{3}.
\end{equation}

Now, in order to compute the effects of the coupling of the excitons
to the LO phonons we will follow a perturbative approach, taking $H_{0}+H_{1}$
as the unperturbed Hamiltonian and $H_{2}+H_{3}$ as the perturbative
term. From second order perturbation theory, we write the energy correction
to the system's ground state as
\begin{equation}
\Delta E=\sum_{\nu_{X}\nu_{ph}}\frac{\Big|\langle1s;n_{ph}(T)|H_{2}+H_{3}|\nu_{X};\nu_{ph}\rangle\Big|^{2}}{E_{GS}-E_{\nu_{X}\nu_{ph}}},\label{eq:Second order pertubation theory}
\end{equation}
where $\nu_{X}$ and $\nu_{ph}$ refer to the states of the exciton
and the phonons, respectively, with a combined energy $E_{\nu_{X}\nu_{ph}}$,
$1s$ refers to the most tightly bound excitonic state and $n_{ph}(T)$
corresponds to the phonon distribution at a temperature $T$, with
a combined energy $E_{GS}$. The sum runs over all the $\nu_{X}$
and $\nu_{ph}$ except $\left\{ \nu_{X},\nu_{ph}\right\} =\left\{ 1s,n_{ph}(T)\right\} .$
A direct evaluation of Eq. (\ref{eq:Second order pertubation theory})
would undoubtedly be a daunting task, with little probability of success,
since all the excitonic wave functions would be required and an infinite
number of matrix elements would have to be evaluated. As an alternative
route, one can follow the Dalgarno-Lewis approach \citep{dalgarno1955exact}
in order to bypass the sum over states. This ingenious approach consists
on the introduction of an operator, defined through a differential
equation, which when inserted in Eq. (\ref{eq:Second order pertubation theory})
allows the sum over states to be removed. The problem of computing
$\Delta E$ is then reduced to the evaluation of a single matrix element.
More specifically, it is possible to show that Eq. (\ref{eq:Second order pertubation theory})
can be written as the sum of four contributions\begin{widetext}
\begin{equation}
\Delta E=\Delta E_{1}+\Delta E_{2}+\Delta E_{3}+\Delta E_{4},
\end{equation}
where
\begin{align}
\Delta E_{1}=\frac{U^{2}}{A}\sum_{\nu_{X}}\sum_{\mathbf{q}}\frac{1}{q}\bigg\{ & \frac{n_{ph}(\mathbf{q},T)+1}{E_{1s}-E_{\nu_{X}}-\hbar\omega_{\mathbf{q}}}\langle1s|e^{-i\mathbf{q}\cdot\left(m_{e}/M\right)\mathbf{r}}|\nu_{X}\rangle\langle\nu_{X}|e^{i\mathbf{q}\cdot\left(m_{e}/M\right)\mathbf{r}}|1s\rangle\nonumber \\
+ & \frac{n_{ph}(\mathbf{q},T)}{E_{1s}-E_{\nu_{X}}+\hbar\omega_{\mathbf{q}}}\langle1s|e^{i\mathbf{q}\cdot\left(m_{e}/M\right)\mathbf{r}}|\nu_{X}\rangle\langle\nu_{X}|e^{-i\mathbf{q}\cdot\left(m_{e}/M\right)\mathbf{r}}|1s\rangle\bigg\},\label{eq:Delta E1}
\end{align}
and
\begin{align}
\Delta E_{3}=-\frac{U^{2}}{A}\sum_{\nu_{X}}\sum_{\mathbf{q}}\frac{1}{q}\bigg\{ & \frac{n_{ph}(\mathbf{q},T)+1}{E_{1s}-E_{\nu_{X}}-\hbar\omega_{\mathbf{q}}}\langle1s|e^{-i\mathbf{q}\cdot\left(m_{e}/M\right)\mathbf{r}}|\nu_{X}\rangle\langle\nu_{X}|e^{-i\mathbf{q}\cdot\left(m_{h}/M\right)\mathbf{r}}|1s\rangle\nonumber \\
+ & \frac{n_{ph}(\mathbf{q},T)}{E_{1s}-E_{\nu_{X}}+\hbar\omega_{\mathbf{q}}}\langle1s|e^{-i\mathbf{q}\cdot\left(m_{h}/M\right)\mathbf{r}}|\nu_{X}\rangle\langle\nu_{X}|e^{-i\mathbf{q}\cdot\left(m_{e}/M\right)\mathbf{r}}|1s\rangle\bigg\},\label{eq:Delta E2}
\end{align}
\end{widetext}with $E_{1s}$ and $E_{\nu_{X}}$ the energies of the
$1s$ and $\nu_{X}$ excitonic states, respectively, and $n_{ph}(\mathbf{q},T)$
the Bose-Einstein distribution function for phonons with energy $\omega_{\mathbf{q}}$
at a temperature $T$. Here we still consider the phonon energy as
a function of the momentum in order to present general expressions,
but later we will consider $\omega_{\mathbf{q}}=\omega_{{\rm LO}}$
when numerical results are computed. The expressions for $\Delta E_{2}$
and $\Delta E_{4}$ follow directly from these two by replacing $m_{e}(m_{h})$
with $m_{h}(m_{e})$. Each of the four contributions is made up of
two terms with distinct physical origin: one originating from phonon
emission and the other from phonon absorption. In the limit of vanishing
temperature only the former contributes due to the process of spontaneous
phonon emission.

As we just mentioned, to forego the sum over states, we will make
use of Dalgarno and Lewis' formulation of perturbation theory. In
order to evaluate $\Delta E_{1}$ we introduce two operators $F_{1\pm}$
which obey to the relations
\begin{equation}
\left(\left[F_{1\pm},H_{0}\right]\pm\hbar\omega_{\mathbf{q}}F_{1\pm}\right)|1s\rangle=e^{\mp i\mathbf{q}\cdot\mathbf{r}m_{e}/M}|1s\rangle.\label{eq:F1pmdef}
\end{equation}
Now, we apply these to Eq. (\ref{eq:Delta E1}), remove the sum over
states and introduce three complete sets of plane waves. Doing so,
and taking advantage of the orthogonality relation between plane waves,
one finds the following expression for $\Delta E_{1}$:\begin{widetext}
\begin{align}
\Delta E_{1}=- & \frac{2\mu}{\hbar^{2}}\frac{U^{2}}{A}\sum_{\mathbf{q}}\sum_{\mathbf{k}}\frac{1}{q}\left[n_{ph}(\mathbf{q},T)+1\right]\frac{\langle1s|\mathbf{k}\rangle\langle\mathbf{k}|1s\rangle}{q^{2}\left(\frac{m_{e}}{M}\right)^{2}+2\mathbf{k}\cdot\mathbf{q}\frac{m_{e}}{M}+2\mu\omega_{\mathbf{q}}/\hbar}\nonumber \\
- & \frac{2\mu}{\hbar^{2}}\frac{U^{2}}{A}\sum_{\mathbf{q}}\sum_{\mathbf{k}}\frac{1}{q}n_{ph}(\mathbf{q},T)\frac{\langle1s|\mathbf{k}\rangle\langle\mathbf{k}|1s\rangle}{q^{2}\left(\frac{m_{e}}{M}\right)^{2}+2\mathbf{k}\cdot\mathbf{q}\frac{m_{e}}{M}-2\mu\omega_{\mathbf{q}}/\hbar}.\label{eq:Delta E1 *}
\end{align}
\end{widetext}Comparing Eqs. (\ref{eq:Delta E1}) and (\ref{eq:Delta E1 *}),
we note that with the approach of Dalgarno and Lewis the problem of
computing $\Delta E_{1}$ was drastically modified. While in Eq. (\ref{eq:Delta E1})
the knowledge of all the excitonic states was required, in (\ref{eq:Delta E1 *})
only the Fourier transform of the $1s$ state wave function is needed.
Moreover, the complexity of the calculations was greatly reduced,
since now we need only compute two sums over the momenta $\mathbf{q}$
and $\mathbf{k}$ while previously, the computation of an infinite
number of matrix elements was required. Finally we note that in Eq.
(\ref{eq:Delta E1 *}) the operators $F_{1\pm}$ are absent, since
we do not need them explicitly, but rather the expression for their
matrix element between plane waves, which can be computed from Eq.
(\ref{eq:F1pmdef}). In order to progress analytically, we follow
a variational path to describe the wave function of the $1s$ excitonic
state. Our variational ansatz, inspired by the Hydrogen atom, reads
\citep{quintela2020colloquium}
\begin{equation}
\psi_{1s}(r)=a\sqrt{\frac{2}{\pi}}e^{-r/a},\qquad\psi_{1s}(\mathbf{k})=\frac{2a\sqrt{2\pi}}{\left(1+a^{2}k^{2}\right)^{3/2}},
\end{equation}
where $a$ is a variational parameter determined from the minimization
of the expectation value of $H_{0}$, and can be roughly interpreted
as the excitonic Bohr radius. This choice of the trial wave function
produces wave functions in good agreement with the ones found using
exact numerical methods. A more sophisticated ansatz combining two
exponential functions \citep{pedersen2016exciton} could also be used.
This option would produce results in perfect agreement with the numerical
ones, with the cost of more involved calculations, without a simple
analytical solution. In any case, the choice of one of the ansätze
over the other should produce only minimal changes in the final result.
To continue with the calculation of $\Delta E_{1}$ the sums over
$\mathbf{q}$ and $\mathbf{k}$ must be converted into two-dimensional
integrals. From this point onward we explicitly consider that $\omega_{\mathbf{q}}=\omega_{{\rm LO}}$;
as a consequence the terms $n_{ph}(\mathbf{q},T)$ become momentum
independent and can be taken out of the integrals. Writing the integrals
in polar coordinates, one finds the following angular integral
\begin{equation}
\int_{0}^{2\pi}\frac{d\theta}{\sigma_{\pm}+k\cos\theta}=\begin{cases}
{\rm sign}\sigma_{\pm}\frac{2\pi}{\sqrt{\sigma_{\pm}^{2}-k^{2}}}, & |\sigma_{\pm}|>1\\
0, & |\sigma_{\pm}|<1
\end{cases}
\end{equation}
where $\sigma_{\pm}=\left(q^{2}\pm2\mu\omega_{{\rm LO}}\right)/2q$.
When $|\sigma_{\pm}|<1$ the principal value of the integral should
be considered. We now note that for the second term in Eq. (\ref{eq:Delta E1 *}),
the term associated with absorption of phonons, $\sigma_{-}=\pm1$
depending on the value of $q$. The same does not apply to the other
contribution, where $\sigma_{+}>0$ $\forall\;q\geq0$. As a consequence
of the two possible signs that originate from the angular integration,
when the integrals over $dk$ and $dq$ are computed the phonon absorption
contribution from $\Delta E_{1}$ (and $\Delta E_{2}$ after the roles
of $m_{e}$ and $m_{h}$ are switched) vanishes. Computing the remaining
integrals, we find
\begin{align}
\Delta E_{1}= & -\frac{4m_{h}}{\pi\hbar^{2}}U^{2}a^{2}\left[n_{ph}(\omega_{{\rm LO}},T)+1\right]\times\\
 & \left[\frac{\pi(\chi_{+}^{2}-1)(4\chi_{+}^{2}+3)}{32a^{3}\beta^{2}\chi_{+}^{3}}+f_{+}\right],
\end{align}
with $\chi_{+}=1+a^{2}\beta^{2},$ $\beta^{2}=2\mu\omega_{{\rm LO}}/\hbar$
and
\begin{equation}
f_{+}=\int_{0}^{\infty}\frac{1}{16q}\frac{3{\rm arcsinh}\left[a\zeta_{+}(q)\right]}{a\left[1+a^{2}\zeta_{+}^{2}(q)\right]^{5/2}}dq\approx\frac{3}{8\pi a},
\end{equation}
where $\zeta_{+}(q)=\left(q^{2}+\beta^{2}\right)/2q$. The value of
$f_{+}$ is roughly less that one half of the the term with which
it is summed. As we have already mentioned, the contribution $\Delta E_{2}$
is obtained from $\Delta E_{1}$ by substituting $m_{h}$ with $m_{e}$.

Now that $\Delta E_{1}$ and $\Delta E_{2}$ were computed, we turn
our focus to the contributions $\Delta E_{3}$ and $\Delta E_{4}$.
The process to compute these terms is very much alike the one described
for the other two. Similarly to before, we start by defining the operators
$F_{3\pm}$ which obey to the relation $\left([F_{3\pm},H]\pm\hbar\omega_{\mathbf{q}}F_{3\pm}\right)|1s\rangle=e^{-i\mathbf{q}\cdot\left(m_{e/h}/M\right)\mathbf{r}}|1s\rangle$.
The introduction of these operators allows us to remove the sum over
states. After the plane wave basis have been introduced, and their
orthogonality relations employed, we arrive at the following expression:\begin{widetext}
\begin{align}
\Delta E_{3} & =-\frac{2\mu}{\hbar^{2}}\frac{U^{2}}{A}\sum_{\mathbf{q}}\sum_{\mathbf{k}}\frac{1}{q}\bigg\{\left[n_{ph}(\mathbf{q},T)+1\right]\frac{\langle1s|\mathbf{k}\rangle\langle\mathbf{k}+\mathbf{q}|1s\rangle}{q^{2}\frac{\left[2m_{e}+m_{h}\right]m_{h}}{M^{2}}+2\mathbf{k}\cdot\mathbf{q}\frac{m_{h}}{M}-2\mu\omega_{\mathbf{q}}/\hbar}\nonumber \\
 & +n_{ph}(\mathbf{q},T)\frac{\langle1s|\mathbf{k}\rangle\langle\mathbf{k}+\mathbf{q}|1s\rangle}{q^{2}\frac{\left[2m_{h}+m_{e}\right]m_{e}}{M^{2}}+2\mathbf{k}\cdot\mathbf{q}\frac{m_{e}}{M}+2\mu\omega_{\mathbf{q}}/\hbar}\bigg\}.\label{eq:Delta E3 *}
\end{align}
\end{widetext}Comparing this result with Eq. (\ref{eq:Delta E1 *}),
we note that in the present case the Fourier transforms of the $1s$
wave function are evaluated at different momenta. This modification
significantly increases the complexity of the integrals that must
be computed, preventing the existence of a simple analytical solution.
The angular integrals can, however, be computed analytically, yielding:
\begin{align}
 & \int_{0}^{2\pi}\frac{d\theta}{\left[\xi_{\pm}+\cos\theta\right]\left[\delta+\cos\theta\right]^{3/2}}=\nonumber \\
= & -4\frac{\left(1+\xi_{\pm}\right)\mathbf{E}\left(\frac{2}{1+\delta}\right)-\left(\delta-1\right)\boldsymbol{\Pi}\left(\frac{2}{1+\xi_{\pm}}\bigg|\frac{2}{1+\delta}\right)}{\left(1+\xi_{\pm}\right)\sqrt{\delta+1}\left(\delta-1\right)\left(\delta-\xi_{\pm}\right)},
\end{align}
where $\mathbf{E}(x)$ and $\boldsymbol{\Pi}(x|y)$ are elliptic integrals
of the second and third kind, respectively, $\delta=\left(a^{-2}+k^{2}+q^{2}\right)/2kq>1$
and $\xi_{\pm}=\left(q^{2}\frac{\left[2m_{h/e}+m_{e/h}\right]}{M}\pm\frac{2m_{h/e}\omega_{{\rm LO}}}{\hbar}\right)/2kq$.
This solution is valid for both $|\xi_{\pm}|>1$ and $|\xi_{\pm}|<1$.
For the latter case, the principal value of the integral should be
considered. The remaining integrals over $dk$ and $dq$ do not yield
analytical solutions, and as a consequence must be evaluated numerically,
taking the principal value of the integral when necessary. When doing
so, one must proceed carefully, starting by determining the points
where poles and branch cuts appear. These points correspond to the
ones where the the conditions $\xi_{-}=-1$ and $\xi_{\pm}=1$ are
satisfied; in addition care must be exercised when $\delta-\xi_{\pm}$.

\begin{figure}[h]
\centering{}\includegraphics{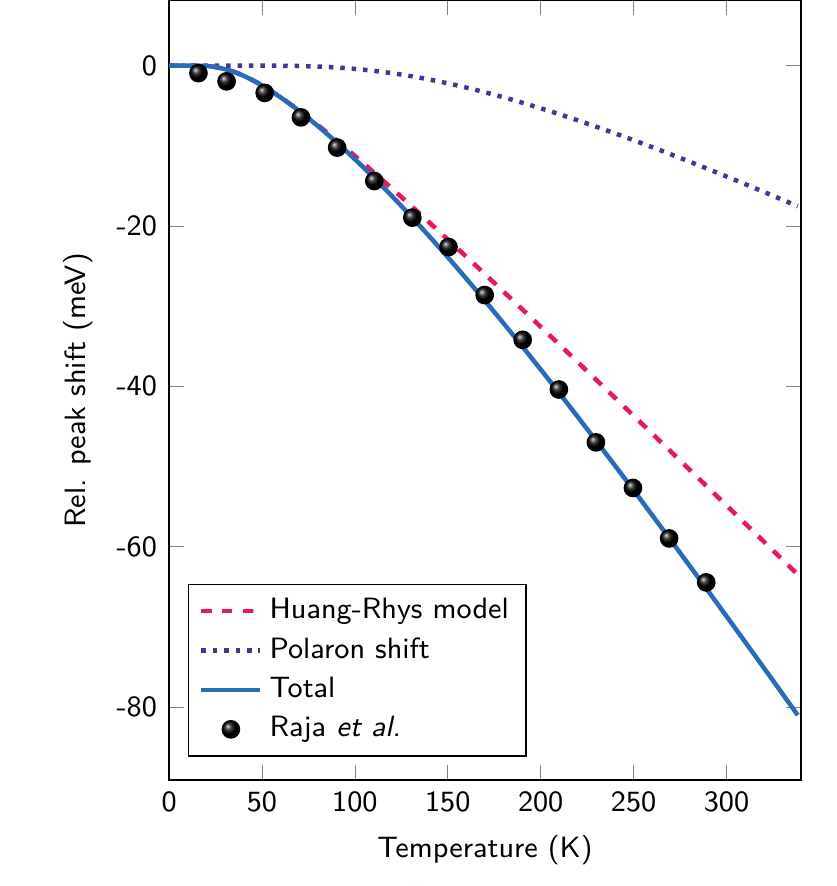}\caption{\label{fig:Fit}Comparison between the fit obtained with Eq. (\ref{eq:Peak shift})
and the experimental data of Ref. \citep{raja2018enhancement}. An
excelent agreement between the theoretical description and the experimental
points is evident. Also depicted are the isolated contributions of
the polaron shift, and the gap renormalization described with the
Huang-Rhys model. The value of the fitting parameters is given in
Table \ref{tab:Parameters}.}
\end{figure}

Up to this point we have described and given equations that characterize
the exciton-phonon interaction. As an application of the results so
far derived we will produce a theoretical description of the experimental
data presented in Ref. \citep{raja2018enhancement}, where the shift
of the $1s$ excitonic resonance was measured as a function of the
temperature. To accurately describe this effect we must consider something
so far ignored. As the temperature increases two distinct effects
take place. On the one hand, the exciton-phonon interaction, dominated
by LO phonons, will modify the exciton binding energy, shifting the
excitonic peak; this is the effect we have theoretically described.
On the other hand, the band gap decreases as the temperature increases,
also contributing to the excitonic resonance shift. To describe this
effect the Varshni empirical model \citep{varshni1967temperature}
is commonly used, however we choose to consider the vibronic model
of Huang and Rhys \citep{huang2000theory}, which takes into account
the effect of acoustic phonons:
\begin{equation}
E_{g}(T)=E_{g}(0)-S\langle\hbar\omega_{{\rm A}}\rangle\left[\coth\frac{\langle\hbar\omega_{{\rm A}}\rangle}{2k_{B}T}-1\right],
\end{equation}
where $E_{g}(T)$ is the band gap magnitude at a temperature $T$,
$\langle\hbar\omega_{{\rm A}}\rangle$ is the mean energy of the acoustic
phonons (about $10$ meV \citep{tongay2012thermally}), $k_{B}$ is
the Boltzmann constant, and $S$ is a fitting parameter of the order
of 1, describing the electron-acoustic-phonon coupling. The expression
for the $1s$ resonance position as a function of $T$, which we label
as$E_{X}(T)$, relatively to its position at $T=0$ K, reads:
\begin{align}
E_{X}(T)-E_{X}(0) & =-S\langle\hbar\omega_{{\rm A}}\rangle\left[\coth\frac{\langle\hbar\omega_{{\rm A}}\rangle}{2k_{B}T}-1\right]\nonumber \\
 & +E_{B}(T)-E_{B}(0),\label{eq:Peak shift}
\end{align}
where $E_{B}(T)<0$ is the $1s$ state binding energy at a temperature
$T$. This quantity can be obtained using \citep{de1973ground}
\begin{equation}
E_{B}(T)=E_{\emph{ab\, initio}}+\Delta E(T)+2\alpha\hbar\omega_{{\rm LO}},
\end{equation}
where $E_{{\rm \textrm{{\normalcolor \ensuremath{\mathit{ab\:initio}}}}}}$
is the $1s$ binding energy of the unperturbed system, that is, when
no phonons are present, $\Delta E(T)=\Delta E_{1}+\Delta E_{2}+\Delta E_{3}+\Delta E_{4}$
and $2\alpha\hbar\omega_{{\rm LO}}$ is the sum of the free electron
and hole polarons (where we assumed that $\alpha$ is approximately
the same for electrons and holes). The value of $E_{{\rm \textrm{{\normalcolor \ensuremath{\mathit{ab\:initio}}}}}}$
can be found numerically or with semi-analytical methods, however,
since this is a temperature independent value, it vanishes from $E_{X}(T)-E_{X}(0)$.
Using Eq. \ref{eq:Peak shift} we performed a fit to the experimental
data of Ref. \citep{raja2018enhancement}. The comparison between
our theoretical description and the experimental results is depicted
in Figure \ref{fig:Fit}; the parameters are presented in Table \ref{tab:Parameters}.
\begin{table}
\centering{}%
\begin{tabular}{ccccccccc}
\toprule 
$\kappa$ & $r_{0}$ & $m_{e}$ & $m_{h}$ & $a$ & $\hbar\omega_{{\rm LO}}$ & $\alpha$ & $\langle\hbar\omega_{{\rm A}}\rangle$ & $S$\tabularnewline
\midrule
\midrule 
2.4 & 37.9 & 0.35 & 0.26 & 15 & 43 & 2 & 11 & 1.32\tabularnewline
\bottomrule
\end{tabular}\caption{\label{tab:Parameters}Parameters used to model the problem, and the
fitting parameters. The distances are given in \AA, the masses in
units of the electron bare mass, and the energies in meV. The value
of $\kappa$ corresponds to the mean dielectric constant of vacuum
and SiO\protect\textsubscript{2}. The value of $r_{0}$ was taken
from Ref. \citep{chaves2017excitonic}, and the values of $m_{e}$
and $m_{h}$ from Ref. \citep{kormanyos2015k}. The value of $a$
was obtained from the minimization of $H_{0}$ using the parameters
just mentioned. The value for the LO phonons energy was taken from
Ref. \citep{tongay2012thermally}. The values of $\alpha$, $\langle\hbar\omega_{{\rm A}}\rangle$
and $S$ were obtained from the fit of the theoretical model to the
data of Ref. \citep{raja2018enhancement}.}
\end{table}
Inspecting Figure \ref{fig:Fit} we observe an excellent agreement
between our theoretical description and the experimental data-points.
At room temperature the gap renormalization is responsible for a shift
of approximately 65meV while the polaron shift contributes with approximately
15meV, in rough agreement with the values found in Ref. \citep{christiansen2017phonon}.
Analyzing the content of Table \ref{tab:Parameters} we note that
the fitting parameters, $\alpha$, $\langle\hbar\omega_{{\rm A}}\rangle$
and $S$ are in agreement with previous results found in the literature.
The value of $\alpha$ lies inside the interval between 0 and 5 indicated
in Ref. \citep{devreese2012physics}. The value of $\langle\hbar\omega_{{\rm A}}\rangle$
matches the one obtained in Ref. \citep{tongay2012thermally} and
used in Ref. \citep{choi2017temperature}, where MoSe\textsubscript{2}
was studied. In Ref. \citep{tongay2012thermally} a value of $S=1.93$
was found for MoSe\textsubscript{2}. Using the fact that this parameter,
which characterizes the coupling to phonons, should be proportional
to the square root of the effective masses, and noting that the effective
masses in MoSe\textsubscript{2} differ from those in WS\textsubscript{2}
approximately by a factor of $0.7^{2}$ \citep{kormanyos2015k}we
can estimate that the value of $S$ in WS\textsubscript{2} should
be around 1.3, in total agreement with the value we found. 

In summary, using second order perturbation theory we have successfully
described the effect of temperature on aset of experimental data-points
on the position of the fundamental absorption line peak of the 1s
excitonic transition (corresponding to the more tightly bound exciton)
in WS$_{2}$ \citep{raja2018enhancement}. The experiment shows a
red shift of the absorption line when the temperature increases. We
were able to describe, in quantitative terms, the observed shift considering
two different effects: the polaron shift and the renormalization of
the single particle gap with temperature. Both effects were shown
to produce a sizable effect to the overall red shift, and should be
accounted for in theoretical descriptions of this phenomenom. We stress
that both effects are due to different set of phonons: the longitudinal
optical phonons in the intrinsic red shift of the absorption line
and the acoustic phonons in the modification of the single particle
gap. A follow up of this work will focus on the calculation of line
shape of the absorption peak as a function of temperature which requires
solving the Bethe-Salpeter equation in the presence of the phonon's
field.

N.M.R.P acknowledges support by the Portuguese Foundation for Science
and Technology (FCT) in the framework of the Strategic Funding UIDB/04650/2020.
J.C.G.H. acknowledges the Center of Physics for a grant funded by
the UIDB/04650/2020 strategic project. N.M.R.P. acknowledges support
from the European Commission through the project ``Graphene-Driven
Revolutions in ICT and Beyond'' (Ref. No. 881603, CORE 3), COMPETE
2020, PORTUGAL 2020, FEDER and the FCT through projects POCI-01-0145-FEDER-028114,
PTDC/NAN-OPT/29265/2017

\bibliographystyle{IEEEtran}

\end{document}